\newcommand{\Ha}{H$\alpha$}
\newcommand{\vel}{$\rm km\ s^{-1}$}
\newcommand{\url}{http://www.pha.jhu.edu/$\sim$danforth/uit/}

\documentstyle[11pt,aaspp4]{article}

\input{psfig}
\begin{document}

\title{A Comparison of Ultraviolet, Optical, and X-Ray Imagery of Selected
Fields in the Cygnus Loop}

\author{\sc Charles W.\ Danforth\altaffilmark{1}, Robert H.\
Cornett\altaffilmark{2}, N. A. Levenson\altaffilmark{1}, William P.\
Blair\altaffilmark{1}, Theodore P.\ Stecher\altaffilmark{3}}

\altaffiltext{1}{Department of Physics and Astronomy, The Johns Hopkins
University, 3400 N. Charles Street, Baltimore, MD 21218; danforth@pha.jhu.edu,
levenson@pha.jhu.edu, wpb@pha.jhu.edu}

\altaffiltext{2}{Raytheon ITSS, 4400 Forbes Blvd., Lanham, MD 20706;
cornett@stars.gsfc.nasa.gov}

\altaffiltext{3}{Laboratory for Astronomy and Solar Physics, NASA/GSFC, Code
681, Greenbelt, MD 20771; stecher@stars.gsfc.nasa.gov}

\begin{center}{Accepted for Publication January 27, 2000}\end{center}

\begin{abstract}
During the Astro-1 and Astro-2 Space Shuttle missions in 1990 and 1995, far
ultraviolet (FUV) images of five 40\arcmin\ diameter fields around the rim of
the Cygnus Loop supernova remnant were observed with the Ultraviolet Imaging
Telescope (UIT).  These fields sampled a broad range of conditions including
both radiative and nonradiative shocks in various geometries and physical
scales.  In these shocks, the UIT B5 band samples predominantly \ion{C}{4}
$\lambda$1550 and the hydrogen two-photon recombination continuum.  Smaller
contributions are made by emission lines of \ion{He}{2} $\lambda$1640 and
\ion{O}{3}] $\lambda$1665.  We present these new FUV images and compare them
with optical \Ha\ and [\ion{O}{3}], and ROSAT HRI X-ray images.  Comparing the
UIT images with those from the other bands provides new insights into the
spatial variations and locations of these different types of emission.  By
comparing against shock model calculations and published FUV spectroscopy at
select locations, we surmise that resonance scattering in the strong FUV
permitted lines is widespread in the Cygnus Loop, especially in the bright
optical filaments typically selected for observation in most previous studies.

\end{abstract}


\section{Introduction}
Because of its large angular size and wide range of shock conditions, the
Cygnus Loop is one of the best laboratories for studying the environment and
physics of middle-aged supernova remnants (SNR).  It covers a huge expanse in
the sky (2.8$\times$3.5$\rm ^o$) corresponding to 21.5$\times$27 pc, at a newly
determined distance of 440 pc (\cite{Blair99}).  The currently accepted view
for the Cygnus Loop is that it represents an explosion in a cavity produced by
a fairly massive  precursor star (cf. \cite{Levenson98}).  The SN shock has
been traveling relatively unimpeded for roughly ten parsecs and has only
recently begun reaching the denser cavity walls.  The size of the cavity
implicates a precursor star of type early B. The interaction of the shock with
the complex edges of the cavity wall is responsible for the complicated mixture
of optical and X-ray emission seen in superposition, and a dazzling variety of
optical filament morphologies.

Portions of the SN blast wave propagating through the fairly rarefied atomic
shell ($<$1 cm$^{-3}$), show faint filaments with hydrogen
Balmer-line-dominated optical spectra.  These filaments represent the position
of the primary blast wave and are often termed nonradiative shocks (because
radiative losses are unimportant to the dynamics of the shock itself).  Ambient
gas is swept up and progressively ionized, emitting \ion{He}{2}, \ion{C}{4},
\ion{N}{5}, and \ion{O}{6} lines in the FUV (Figure~1, bottom spectrum)
(\cite{Hester94}, \cite{Raymond83}).  Balmer-dominated emission arises from the
fraction ($\sim$0.3) of neutral hydrogen swept up by the shock that stands some
chance of being excited and recombining before it is ionized in the post-shock
flow (\cite{Chevalier78}; \cite{CKR80}).

The Balmer emission is accompanied by hydrogen two-photon events which produce
a broad continuum above 1216\AA\ peaking at $\sim$1420\AA\
(\cite{Nussbaumer84}).  For recombination and for high temperature shocks, the
ratio of two-photon emission to Balmer  is nearly constant ($\sim$8:1).  In
slow shocks ($\sim$40\vel) in neutral gas, the ratio can be enhanced
considerably (\cite{Dopita82}).

Balmer-dominated filaments are very smooth and WFPC2 observations by Blair et
al. (1999) show that they are exceedingly thin as well---less than one WFC
pixel across when seen edge-on, or $<6 \times 10^{14}$  cm at our assumed
distance, in keeping with theoretical predictions (cf. \cite{Raymond83}).
Postshock temperatures reach millions of degrees and the hot material emits
copious soft X-rays.  The density is low, however, and cooling is very
inefficient.  With time, as the shock continues to sweep up material, these
filaments will be able to start cooling more effectively and will evolve to
become radiative filaments.

The bright optical filaments in the Cygnus Loop represent radiative shocks in
much denser material, such as might be expected in the denser portions of the
cavity wall.  These shocks are said to be radiative (that is, energy losses
from radiation are significant); they have more highly developed cooling and/or
recombination zones. The shocked material emits in the lines of a broad range
of hot, intermediate,  and low temperature ions, depending on the effective
`age' of the shock at a given location and the local physical conditions.  For
instance, a relatively recent encounter between the shock and a density
enhancement (or similarly, a shock that has swept up a fairly low total column
of material) may show very strong [O~III] $\lambda 5007$ compared with \Ha.
This would indicate that the coolest part of the flow, the recombination zone
where the Balmer lines become strong, has not yet formed.  Such shocks are said
to be `incomplete' as the shocked material remains hot and does not yet emit in
the lower ionization lines.

In contrast, radiative filaments with the full range of ionization (including
the low ionization lines) are well approximated by full, steady-flow shock
model calculations, such as those of \cite{Raymond79}, \cite{Dopita84}, and
Hartigan, Raymond, \& Hartmann (1987; hereafter \cite{HRH}).  Morphologically,
radiative complete filaments lack the smooth grace of nonradiative  filaments
or even radiative incomplete filaments in some cases (cf. \cite{Fesen82}).  The
more irregular appearance of these filaments is due  partly to inhomogeneities
in the shocked clouds themselves, partly to  turbulence  and/or thermal
instabilities that set in during cooling (cf. \cite{Innes92} and references
therein), and partly to several clouds appearing along single  lines of sight.
Often the emission at a given filament position cannot be  characterized by a
single shock velocity.

Much of the above understanding of shock types and evolutionary stages has been
predicated on UV/optical studies of the Cygnus Loop itself.  The Cygnus Loop is
 a veritable laboratory for such studies because of its relative proximity,
large angular extent and low foreground extinction (E[B $-$ V] = 0.08;
\cite{Fesen82}), and thus its accessibility across the electromagnetic
spectrum.  However, because of the range of shock interactions and shock types,
coupled with the significant complication of projection effects near the limb
of the SNR, great care must be taken in order to obtain a full understanding of
what is happening at any given position in the nebular structure.

Although FUV spectra are available at a number of individual filament locations
from years of observations with IUE and the shuttle-borne Hopkins Ultraviolet
Telescope (HUT), the perspective obtainable from FUV imaging has been largely
lacking.  The Ultraviolet Imaging Telescope (UIT) was flown as part of the
Astro-1 Space Shuttle mission in 1990 was used to observe a field in the Cygnus
Loop through both mid-UV and far-UV  (FUV) filters (\cite{Cornett92}).  In this
paper, we report on additional FUV observations with UIT obtained during the
Astro-2 shuttle mission in 1995.  In addition to the field imaged during
Astro-1, UIT observed four different regions around the periphery of the Cygnus
Loop with a resolution comparable to existing optical and X-ray observations.
These fields sample the full range of physical and shock conditions and
evolutionary stages in the SNR.  We combine these data with existing
ground-based optical images and ROSAT HRI X-ray data to obtain new insights
into this prototypical SNR and its interaction with its surroundings.

In \S2 we present the observations obtained with the UIT and review the
comparison data sets.  In \S3 we discuss the spectral content of the UIT filter
used in  the observations.  In \S4 we discuss examples of the various kinds of
shocks as seen in the UIT fields, and summarize our conclusions in \S5.

\section{UIT Observations and Comparison Data}

UIT has flown twice on the Space Shuttle as part of the Astro-1 and Astro-2
programs (1990 December 2-10 and 1995 March 2-18).  Together with the Hopkins
Ultraviolet Telescope (HUT) and the Wisconsin Ultraviolet Photo-Polarimeter
Experiment, UIT explored selected UV targets.  An f/9 Ritchey-Chretien
telescope with a 38 cm aperture and image intensifier systems produced images
of circular 40\arcmin\ fields of view with $\sim$3\arcsec\ resolution at field
center (depending on pointing stability).  Images were recorded on 70mm Eastman
Kodak IIa-O film which was developed and digitized at NASA/GSFC and processed
into uniform data products.  Technical details on the hardware and data
processing can be found in \cite{Stecher92} and Stecher et al. (1997).

\noindent Table~1: UIT B5 Filter Observations in the Cygnus Loop\\
\begin{tabular}{lllll}
Position     & RA(J2000)& Dec(J2000)& exposure (sec) & Figure\\
\hline
W cloud      & 20:45:38 & +31:06:33 & 1010 & 3\\
NE nonrad    & 20:54:39 & +32:17:29 & 2041 & 4\\
NE cloud     & 20:56:16 & +31:44:34 &  500 & 5\\
XA region$\rm^a$& 20:57:35 & +31:07:28 & 1280 & 6\\
XA region    & 20:57:04 & +31:07:45 & 1151 & 6\\
XA region    & 20:57:22 & +31:04:02 & 1516 & 6\\
XA region    & 20:57:24 & +31:03:51 & 1274 & 6\\
SE cloud     & 20:56:05 & +30:44:01 & 2180 & 7\\
\hline
\end{tabular}\\
$\rm^a$ Astro-1 image (cf. \cite{Cornett92})

Astro/UIT images are among the few examples of FUV images of SNRs, and UIT's B5
bandpass ($\sim1450$\AA\ to $\sim1800$\AA) encompasses severally generally
high-excitation and heretofore unmapped lines that are often present in SNR
shocks (Figure~1).  UIT's two Astro flights have produced eight FUV images of
five different Cygnus Loop fields.  Table~1 lists the observation parameters
and field locations, which are indicated in Figure~2.  We will refer to these
fields by the names listed in Table~1.  Since all four exposures of the XA
region (named by \cite{HesterCox86}) are reasonably deep, we constructed a
mosaic of the field using the IRAF\footnotemark\ IMCOMBINE task, resulting in
significantly improved signal-to-noise in the overlapped region of the combined
image. In panel c of Figures 3 through 7, we show the five reduced UIT images
as observed in the B5 filter bandpass.  \footnotetext{IRAF is distributed by
the National Optical Astronomy Observatories, which is operated by the
Association of Universities for Research in Astronomy, Inc.\ (AURA) under
cooperative agreement with the National Science Foundation.}

UIT images with long exposure times suffer from an instrumental malady dubbed
``measles'' by the UIT team (\cite{Stecher97}).  Measles manifest themselves as
fixed-pattern noise spikes in images with a large sunlight flux, such as long
daylight exposures or images of red, very bright sources (e.g. planets or the
Moon).  This effect is probably produced by visible light passing through
pinholes in either the output phosphor of the first stage or the bialkali
photocathode of the second stage of the UIT FUV image tube.  The Cygnus Loop
was a daytime object for both the Astro-1 and Astro-2 flights, but the
phenomenon is visible only in some of the longer exposures.  Most dramatically,
measles are seen in the northeast cloud nonradiative image (Figure~4) as a
darkening in the northwest corner; the individual ``measles'' are spread into a
background by the binning used to produce these images.  Various approaches to
removing the appearance of measles were attempted but none of them have yielded
satisfactory results.  In practice, the measles, here arising from daylight sky
contamination, affect our analysis only by adding to the background level, so
the original images are presented here, ``measles'' and all.

For comparison with our FUV images, we show narrow-band optical images in
[\ion{O}{3}] $\lambda$5007 and \Ha+[\ion{N}{2}] (which for simplicity we refer
to as \Ha) obtained with the Prime Focus Corrector on the 0.8 m telescope at
McDonald Observatory (cf. \cite{Levenson98}).  These images, shown in panels~a
and b of Figures~3~--~7, are aligned and placed on a common scale of 5\arcsec\
per pixel, which is similar to the FUV resolution of 3\arcsec. The optical
images have each been processed with a 3-pixel median filter to remove faint
stars and stellar residuals.

In addition, we show the soft X-ray (0.1--2.4 keV) emission for each field, as
observed with the {\it ROSAT} High Resolution Imager (HRI) (from
\cite{Levenson97}).  The resolution of the HRI imager is 6\arcsec\ on axis
degrading to 30\arcsec\ at the edge of each field.  As with the optical data,
the X-ray images are aligned on a 5\arcsec\ per pixel scale.  The HRI images
have additionally been smoothed with a 3-pixel FWHM gaussian and are shown in
panel d of Figures~3~--~7.  All images in Figures~3~--~7 are displayed on a
logarithmic scale.

Figure~8 shows three-color composite images using \Ha\ as red, B5 as green, and
the {\it ROSAT} HRI as blue.  The color levels have been adjusted for visual
appearance, to best show the relative spatial relationships of the different
emissions.  (The color composite for the Northeast nonradiative region is not
shown, since little new information is gained above Figure~4 and because of the
adverse effect of the measles.)  This will be discussed in more detail below.

\section{Spectral Content of UIT Images}

Figure~1 shows the UIT B5 filter profile superimposed on spectra of typical
radiative and nonradiative filaments, as observed by HUT.  Unlike typical SNR
narrow band images in the optical, the B5 filter is relatively broad and does
not isolate a single spectral line, but rather encompasses several strong,
moderately high ionization lines that are variable from filament to filament.
Cornett et al. (1992) point out that \ion{C}{4} should dominate emission in
this bandpass since it is a strong line centered near the filter's peak
throughput, and since shock models predict this result for a range of important
velocities (cf. Figure~10 and accompanying discussion).  Here we look at this
more closely over a larger range of shock  velocities, and in particular also
discuss the potential complicating effects of hydrogen two-photon recombination
continuum emission, shock completeness, and resonance line scattering.

Empirical comparisons of IUE and HUT emission line observations can be used to
quantify at what level the C~IV emission is expected to dominate the line
emission detected through the B5 filter. For instance, in the highly radiative
XA region (see Figure~7) we have compared a large number of FUV  spectra both
on and adjacent to bright optical filaments against the throughput curve of B5
(Danforth, Blair, \& Raymond 2000; henceforth \cite{DBR}). This comparison
shows that on average the various lines contribute as follows:
\ion{C}{4} $\lambda$1550, 42\%;
\ion{O}{3}] $\lambda$1665, 27\%;
\ion{He}{2} $\lambda$1640, 17\%;
\ion{N}{4} $\lambda$1486, 8\%;
and 6\%
from fainter emission lines.  Using the HUT observation of \cite{Long92}, we
estimate for nonradiative shocks the B5 contributions are more like
\ion{C}{4} (60\%),
\ion{He}{2} (28\%),
and all other species 12\%.
These percentages are only approximate, of course, and will vary with shock
velocity, geometry and a host of other conditions, but they serve to highlight
the fact that, while \ion{C}{4} is the strongest contributor to the line
emission, it is not the only contributor.

In addition, while it is not obvious at the scale of Figure~1, a low level
continuum is often seen in IUE and HUT spectra of Cygnus Loop filaments,
especially where optical \Ha\ emission is present and strong.  This continuum
arises due to the hydrogen two-photon process (cf. \cite{Osterbrock89}).
\cite{Benvenuti80} note that SNR shocks cause two-photon emission from hydrogen
via both collisional excitation and recombination into the 2$^2$S$_{1/2}$
state.  The two-photon spectrum arises from a probability distribution of
photons that is symmetric about 1/2 the energy of Ly$\alpha$ (corresponding to
2431\AA), resulting in a shallow spectral peak near 1420\AA\ and extending from
1216\AA\ towards longer wavelengths, throughout the UV and optical region.  The
expected (integrated) strength of this component is about 8 $\times$ the \Ha\
flux but is spread over thousands of Angstroms.  However, the wide bandpass of
the B5 filter detects $\sim$15\%
of the total two-photon flux available, enough to compete with line emission in
the bandpass.  Further complicating the question, two-photon emission can also
be highly variable from filament to filament.

By using signatures from the images and spectra at other wavelengths, we can
interpret, at least qualitatively, what is being seen in the UIT images. For
instance, Figure~4 shows the NE rim of the SNR.  The faint, smooth \Ha\
filament running along the edge of the X-ray emission is clearly a nonradiative
filament.  The faint emission seen in the B5 image traces these faint Balmer
filaments well, and at this position, does not correlate particularly well with
the clumpy [\ion{O}{3}] emission seen near the middle of the field.  This
implies a relatively strong contribution from two-photon continuum, although as
shown in the bottom spectrum of Figure~1, \ion{C}{4} and \ion{He}{2} are also
present in the filaments at some level.

As discussed earlier, in radiative filaments, higher ionization lines such as
\ion{O}{6} $\lambda$1035, \ion{N}{5} $\lambda$1240, \ion{C}{4} $\lambda$1550,
and [\ion{O}{3}] $\lambda$5007 become  strong first, followed by lower
ionization lines like [\ion{S}{2}] $\lambda$6725, [\ion{O}{1}] $\lambda$6300,
and the hydrogen Balmer lines.  Hence, in filaments that show high optical
[\ion{O}{3}] to \Ha\ ratios, and are thus incomplete shocks, the B5 content
primarily arises from \ion{C}{4} and other line emission.  In older, more
complete shocks where the optical [\ion{O}{3}] to \Ha\ ratios are close to
those expected from steady flow shock models, two-photon emission again should
compete with the line emission and the B5 flux should arise from both sources.
It is difficult to assess these competing effects from Figures 3 -- 7 since the
relative intensities of the two optical images are not always obvious, but much
of the variation in coloration in Figure~8 for bright radiative filaments is
due to the variation in relative amounts of line  emission and two-photon
continuum contributions to the B5 image.

In Figure~9, we show the XA field as seen with UIT (panel a) and ratio maps of
the UIT image against the aligned optical \Ha\ and [\ion{O}{3}] images.  Since
the ionization energies of \ion{C}{4} (64.5 eV) and \ion{O}{3} (54.9 eV) are
similar  (and to the extent that the B5 image contains a substantial component
of \ion{C}{4} emission), we would expect a ratio of B5 to \Ha\ to show evidence
for the transition from incomplete to complete shock filaments.  Such a ratio
map is shown in panel b of Figure~9, and a systematic pattern is indeed seen.
The white filaments, indicative of a relatively low value of the ratio (and
hence relatively strong \Ha\ filaments) tend to lie systematically to the
right.  These filaments tend to be closer to the center of the SNR, and hence
should have had more time (on average) to cool and recombine.  Of course, there
is significant evidence for projection effects in this complicated field as
well.  Indeed, one interpretation of Figure~9b is that we are separating some
of these projection effects, and are seeing two separate `systems' of filaments
that are at differing stages of completeness.

Another way of assessing the expected contributions of line emission and
two-photon emission to the B5 flux is by comparing to shock model calculations.
We use the equilibrium preionization ``E'' series shock models of Hartigan,
Raymond, \& Hartmann (1987, \cite{HRH}) to investigate variations in spectral
contributions to the UIT images as a function of shock velocity.  Figure~10
shows how various spectral components are predicted to change in relative
intensity as shock velocity increases for this set of planar, complete, steady
flow shock models.  As expected, the key contributors to the B5 bandpass are
indeed \ion{C}{4} and two-photon continuum, although between $\sim$100 --
200~\vel\ these models indicate \ion{C}{4} should dominate.

This is quite at odds with `ground truth', as supplied by careful comparisons
at the specific locations of IUE and HUT spectra within the UIT fields of
view.  We note that the two-photon flux per \AA\ in HUT and IUE spectra is  low
and thus difficult to measure accurately since background levels are  poorly
known.  Even so, it is  quite clear from comparisons such as those of
\cite{Benvenuti80} and Raymond et al. (1988) that nowhere do we see \ion{C}{4}
dominate at the level implied by Figure~10.  (Indeed such studies indicate that
two-photon should dominate!  As will be discussed more thoroughly in \S4,
\cite{Benvenuti80} and others give two-photon fluxes which overwhelm \ion{C}{4}
in the B5 band by a factor of 5-10.  Interestingly, consideration of
incompleteness effects only serves to exacerbate this discrepancy since the
expected two-photon emission should be weaker or absent.  Something else is
going on.

That `something else' is apparently resonance line scattering.  It has long
been suspected that the strong UV resonance lines, like \ion{N}{5}
$\lambda$1240, \ion{C}{2} $\lambda$1335, and \ion{C}{4} $\lambda$1550, are
affected by self-absorption along the line of sight, either by local gas within
the SNR itself or by the intervening interstellar medium.  We can expect
significant column depth from the cavity wall of the remnant itself.  Since
filaments selected for optical/UV observation have tended to be bright, and
since many such filaments are edge-on sheets of gas with correspondingly high
line of sight column densities (\cite{Hester87}), the spectral observations are
likely affected in a systematic way.

While this has been known for some years (\cite{Raymond81}), the UIT data
presented here indicate just how widespread resonance line scattering is in the
Cygnus Loop and how  significantly the \ion{C}{4} intensity may be reduced by
this effect.  Figure~9c  shows a ratio map of the B5 image to the [\ion{O}{3}]
optical image of the XA region  (cf. \cite{Cornett92}).  Since [\ion{O}{3}] is
a forbidden transition, its optically thin emission is not  affected by
resonance scattering.  The ionization potentials for \ion{C}{4} and
[\ion{O}{3}] are similar, so this ratio should provide some information about
resonance scattering, if a significant fraction of the B5 image can be
attributed to \ion{C}{4}. Hence, this ratio image shows where resonance line
scattering is most  important, and provides information on the 3-dimensional
structure of  regions within the SNR.

The B5 image gives the {\em appearance} of smaller dynamic range and lower
spatial  resolution than [O~III] because we see optically thick radiation from
only a  short distance into the filaments.  The highest saturation (lowest
ratios, or light areas in Figure~9c) occurs in the cores of filaments and dense
clouds, such as the three regions indicated in Figure~9a.  The ``spur''
filament was studied in detail by \cite{Raymond88} and is probably an edge-on
sheet of gas.  The region marked `B' is the turbulent, incomplete shocked cloud
observed with HUT during Astro-1 (\cite{Blair91}).  The XA region is also a
shocked cloud or finger of  dense gas that is likely elongated in our line of
sight (cf. \cite{HesterCox86}; \cite{DBR}). What is surprising, however, is the
extent to which the light  regions in Figure~9c extend beyond the cloud cores
into regions of more  diffuse emission.  This indicates that significant
resonance scattering is very  widespread in the Cygnus Loop. The diminished
\ion{C}{4} flux also boosts the  relative importance of two-photon emission in
the B5 bandpass and explains  the discrepancy between numerous spectral
observations and the shock model  predictions shown in Figure~10.

UIT's B5 images are particularly useful in that they sample two important shock
physics regimes--the brightest radiative shocks arising in dense clouds and the
primary blast wave at the edge of the shell.  However, it is evidently
difficult to predict the spectral content of B5 images alone without detailed
knowledge of the physics of the emitting regions.  Nonetheless, B5 images are
useful in combination with [\ion{O}{3}]$\lambda$5007 and \Ha\ images as
empirical tools. The image combinations allow us to determine whether
\ion{C}{4} or two-photon dominates, in two clear-cut cases.  1) In regions
where B5 images closely resemble [\ion{O}{3}] images, the B5 filter is
detecting radiative shocks with velocities in the range 100-200 \vel\ and
therefore primarily \ion{C}{4}.  2) In regions where B5 images closely resemble
\Ha, the B5 filter is detecting largely two-photon emission from recombination
of hydrogen in radiative shocks or from collisional excitiation of hydrogen in
nonradiative shocks.

\section{Discussion}

Each of the fields in our study portrays a range of physical conditions  and
geometries, and hence filament types, seen in projection in many cases.  By
comparing the UV, X-ray and optical emissions, we can gain new insights into
these complexities.  In this section, we discuss the spatial relationships
between the hot, intermediate and cooler components seen in these images.

\subsection{The Western Cloud}

In the Western Cloud field (Figure~3) the B5 image of the bright north-south
filaments resembles the [\ion{O}{3}]5007\AA~ images very closely.  The
filaments are clearly portions of a radiative shock viewed edge-on to our line
of sight.  The Western Cloud has been studied spectroscopically at optical
wavelengths by \cite{Miller74} and in the FUV by \cite{Raymond80b}.

This region shows a case where a cloud is evidently being overrun by a shock,
and the cloud is much larger than the scale of the shock.  The cloud is
elongated in the plane of the sky of dimensions perhaps 1$\times$10 pc
(\cite{Levenson98}) and represents an interaction roughly 1000 years old
(\cite{Levenson96}).  The main north-south radiative filament is bright in all
wavelengths, with good detailed correlation between B5 and [\ion{O}{3}].  \Ha\
is seen to extend farther to the east, toward the center or 'behind' the shock,
as is expected in a complete shock stratification.

Bright X-rays (Figure~3d) are seen to lag behind the radiative filaments by
1-2\arcmin\ (0.15 to 0.3 pc).  This is indicative of a reverse shock being
driven back into the interior material from the dense cloud.  This
doubly-shocked material shows enhanced brightness of about a factor of 2.  From
this, Levenson et al. (1996) derive a cloud/ambient density contrast of about
10.

Attempts to fit shock models to optical observations of the bright filament
have been frustrated by the large [\ion{O}{3}]/H$\beta$ ratio.  A shock
velocity of 130 \vel\ was found by \cite{Raymond80b} using  IUE line strengths
and assuming a slight departure from steady flow and depleted abundances in
both C and Si.  \cite{Raymond80b} also note that much of the hydrogen
recombination zone predicted by steady flow models is absent, implying that the
interaction is fairly young.

As seen in the \Ha\ image, a Balmer-dominated filament projects from the south
of the bright radiative filament toward the northwest. \cite{Raymond80a} find
that the optical spectrum of the filament contains nothing but hydrogen Balmer
lines.  High-resolution observations of the \Ha\ line (\cite{Treffers81}) show
a broad component and a narrow component, corresponding to the pre- and
post-shock conditions in the filament, with a resulting estimated shock
velocity of 130-170 \vel.  The filament may be a foreground or background piece
of the blast wave not related to the radiative portion of the shock, or a
related piece of blast wave that is travelling through the atomic (rather than
molecular) component.  It is visible in both \Ha\ (Figure~3a) and B5
(Figure~3c) though generally not in other bands; thus the B5 flux for this
filament arises primarily from the two-photon process.  There is a small
segment of the filament visible in [\ion{O}{3}] where the shock may be becoming
radiative, visible in B5 as a brightening near the southern end of the
filament.

The X-ray luminosity behind this nonradiative filament is much fainter than
that observed to the east of the main radiative filament, since there is no
reverse shock associated with the nonradiative filament to boost the brightness
(\cite{Hester94}).  The absence of X-rays to the west of this filament confirms
that it represents the actual blast front.  As expected, the peak X-ray flux
lags behind the \Ha\ and B5 flux by roughly one arcminute (0.1 pc).  This
representing the ``heating time'' of gas behind the shock.

A CO cloud is seen just to the south of the Western Cloud field
(\cite{Scoville77}).  The presence of CO clearly indicates material with
molecular hydrogen at densities of 300-1000 cm$^{-3}$.  The nonradiative
filament runs closely along the T$_{antenna}$=5K contour of the CO cloud,
indicating this shock is moving through the atomic component at this stage, but
showing no sign of interaction with the molecular cloud.

\subsection{Northeast Nonradiative Region}

The canonical example of nonradiative filaments in any context lies on the
north and northeast rim of the Cygnus Loop.  There, smooth Balmer filaments
extend counterclockwise from the northern limb (Figure~4), and can be seen
prominently in \Ha\ in Figure~5a.  Small portions of this shock system have
been extensively studied by Raymond et al. (1983), Blair et al. (1991),
\cite{Long92}, Hester, Raymond \& Blair (1994), and most recently by Blair et
al. (1999).  The filaments are clearly visible in \Ha\ (Figure~4a) as well as
B5 (Figure~4c), but invisible  along most of their length in [\ion{O}{3}]
(Figure~4b) except for small segments. These segments represent portions of the
shock front where a slightly higher density has allowed the shock to become
partially radiative.  The shocked, T$\sim10^{6}$K gas emits in an
edge-brightened band of X-rays (Figure~4d).  The brightness variations in
X-rays confirm that the nonradiative filaments are simply wrinkles in the blast
wave  presenting larger column densities to our line of sight.

Spectroscopic observations of selected locations on the filaments indicate that
the B5 filter observes nonradiative filaments as a mixture of \ion{C}{4} and
two-photon emission.  \cite{Long92} find an intrinsic ratio of two-photon
emission to \ion{C}{4} of 4.3, which gives an observed ratio in B5 of 0.65.
Raymond et al. (1983) find fluxes in the same filament which give an observed
ratio of 1.6; in a nearby filament, Hester, Raymond \& Blair (1994) find a
ratio near 2.0.  These filaments all have velocities of around 170 \vel.  It is
likely that much of the ISM carbon is locked up in grains in the preshock
medium, thus boosting the ratio.

The system of thin filaments in the NE nonradiative field extends to the south
and is visible in \Ha\ ahead of the radiative Northeast Cloud (Figure~5)
discussed below.

\subsection{The Northeast Cloud}

The Northeast Cloud (Figure~5) radiative filaments, south and east of the field
discussed above, make up one of the brightest systems in the Cygnus Loop.  The
interaction of the SN blast wave and the denser cavity wall is most evident at
this location.  A complex of radiative filaments can be seen, apparently
jumbled together along our line of sight, displaying the signs of a complete
shock undergoing radiative cooling.  The X-ray edge marking the SN blast wave
is well separated from the  optical and UV filaments, implying a strongly
decelerated shock and cooling that has continued for some time.  Stratification
of  different ionic species is evident, with [\ion{O}{3}] in sharp filaments to
the  east, and more diffuse \Ha\ behind (Figure~8b).

The Northeast Cloud extends into the southern portions of the NE nonradiative
field (Figure~4) as well.  However, the exposure time  for this FUV image is a
factor of four shorter than that in Figure~4c, so the nonradiative filaments
are not detected above the background. There are a few UV-bright sections which
correspond closely with bright [\ion{O}{3}] knots.  However, other equally
bright [\ion{O}{3}] knots in the region do not have corresponding FUV knots.
This may be evidence for a range of shock velocities, or it may be portions of
the shocks that are in transition from nonradiative to radiative conditions.

Using IUE spectra \cite{Benvenuti80} measure the two-photon continuum for one
of the brightest radiative positions within the NE cloud, with a resulting
observed two-photon/\ion{C}{4} ratio of 5.0.  Observations of other radiative
regions both in the Cygnus Loop and in other SNRs similar in morphology and
spectrum give ratios between 1.7 and 10 (\cite{Raymond88}; various unpublished
data).  Therefore, while conditions vary widely within these shocked regions,
spectroscopy indicates that resonance scattering of \ion{C}{4} causes us to see
2-6 times more flux from two-photon emission than from other ions in the field.
 Yet the B5 morphology of most of the field resembles [\ion{O}{3}] far more
than \Ha, as we would expect if two-photon emission were dominant.  The
apparent conflict is likely caused by the fact that most lines of sight through
this region undoubtedly encounter material with a broad range of physical
conditions.  Furthermore, the UIT NE cloud exposure is the shortest of our set.
 Only regions bright in both \Ha\ and [\ion{O}{3}] show up in B5.

\subsection{The XA Field}

The XA field (Figure~6) is a complicated region of predominantly radiative
filaments, noteworthy because an extremely bright and sharp X-ray edge
corresponds closely to a bright knot of UV/visible emission
(\cite{HesterCox86}).  Indeed, this region is seen to be bright in many
wavelengths including radio (\cite{Green90}; \cite{Leahy97}) and infrared
(\cite{Arendt92}).  Strong \ion{O}{6} $\lambda$1035 emission is seen
(\cite{Blair91}) as well as other high-ionization species; \ion{N}{5},
\ion{C}{4}, \ion{O}{3}] (\cite{DBR}) and [\ion{Ne}{5}] (\cite{Szentgyorgyi99}).
See DBR for a more detailed analysis of this region.

In general, the B5 emission corresponds closely to optical [\ion{O}{3}].
However, while optical images show a high contrast between the brightest
`cloud' regions and others in 'empty' space, B5 contrast is lower
(\cite{Cornett92}).  This suggests contributions from a high column depth of
diffuse \ion{C}{4} and/or two-photon emission.  We are either looking at
diffuse material through the edge of a cavity wall or are seeing emission from
face on sheets of gas.  \cite{DBR} show evidence that the bright 'cloud' in the
center is not isolated and may be a density enhancement in the cavity wall or a
finger of denser material projecting in from the east.  The entire blast wave
in the region appears indented from the otherwise circular extent of the SNR
(\cite{Levenson97}) implying that the disturbance is produced by a cloud
extended several parsecs in our line of sight.  The visible structure is likely
the tip of a much larger cloud.

Levenson et al. (1998) suggest a density enhancement in the cavity wall,
resulting in rapid shock deceleration and accounting for the bright emission.
IUE and HUT observations show evidence for a 150 \vel\ cloud shock in the dense
core of XA itself (the west-pointing V shape in the center of the field) and a
faster, incomplete shock in the more diffuse regions to the north and south
(\cite{DBR}).  Two parallel, largely east-west filaments are seen flanking the
central 'cloud'.  The X-ray emission is seen to drop off dramatically south of
the two long radial filament systems.

Blair et al. (1991) report HUT observations of a radiative but incomplete cloud
shock directly to the north of XA marked `B' in Figure~9a.  This region
features almost complete cooling with the exception of \Ha\ and cooler ions.
Raymond et al. (1988) studied the Spur filament and found a completeness
gradient along the length of it.  This filament is well-defined in B5 as well
as the optical bands.

The XA region is the one region in the Cygnus Loop where preionization is
visible ahead of the shock front (\cite{Levenson98}).  This preionization is
caused by X-ray flux from the hot, postshock gas ionizing neutral material
across the shock front.  The emission measure is high enough in this
photoionized preshock gas that it is clearly visible as a diffuse patch of
emission a few arc minutes to the east of the main XA knot in the center of the
field in both \Ha\ and B5.  The B5 flux presumably arises almost entirely from
two-photon emission in this case since no [\ion{O}{3}] is seen (and hence no
strong UV line emission is expected).

One unique ability of the B5 filter becomes apparent in the XA region; that of
detecting nonradiative shocks in ionized gas.  In the X-ray (Figure~6d) we see
a bulge of emission to the north and east of the brightest knot (Hester \&
Cox's XA region proper).  This bulge does not show up in either of the optical
bands, but the perimeter is visible in the FUV at the edge of the X-ray
emission in Figure~6c.  This region has likely been ionized by X-ray flux from
the hot post-shock gas.  A nonradiative shock is now propagating through it
and, lacking a neutral fraction to radiate in \Ha, is seen only in high ions
such as \ion{C}{4}.  This filament is becoming more complete in its southern
extremity (the `B' location in Figure~9a) and is emitting in [\ion{O}{3}] as
well.  This filament also appears to connect to the nonradiative filament seen
in \Ha\ in the Northeast cloud (Figure~5a).

\subsection{The Southeast Cloud}

The Southeast Cloud (Figure~7) presents an interesting quandry.  In the optical
it appears as a small patch of radiative emission with a few associated
nonradiative filaments.  \cite{Fesen92} hypothesize that it represents a small,
isolated cloud at a late stage of shock interaction.  Indeed, the resemblance
to the late-stage numerical models of \cite{Bedogni90} and \cite{Stone92} is
striking.

More recent X-ray analysis (\cite{Graham95}) suggests that the shocked portion
of the southeast cloud is merely the tip of a much larger structure.  Indeed,
it is probably similar to the Western and Northeastern Clouds but at an even
earlier point in its evolution.  \cite{Fesen92} note that the age of the
interaction is probably $4.1\times 10^3$ years based on an assumed blast wave
velocity.  Given the revised distance estimate of Blair et al. (1999), this age
becomes $2.3\times 10^3$ years.

In \Ha\ (Figure~7a) we see a set of nonradiative filaments to the southeast of
the cloud.  These filaments are visible very faintly in B5 (Figure~7c) as
well.Given the complete lack of X-ray emission (Figure~7d) to the east, these
filaments are the primary blast wave.  The fact that these filaments are
indented from the circular rim of the SNR implies the blast wave is diffracting
around some object much larger than the visible emission and extended along our
line of sight (\cite{Graham95}).

Fesen et al. identify a filament segment seen to the west of the SE
cloud--visible in both \Ha\ and our B5 image--as a reverse shock driven back
into the shocked medium.  The X-ray emission, however, demonstrates that this
is instead due the primary forward-moving blast wave.  X-ray enhancement is
seen to the west of the cloud, not the east as we would expect from a doubly
shocked system.  Furthermore, the optical filament is Balmer-dominated, which
requires a significant neutral fraction in the pre-shock gas, which would not
occur at X-ray producing temperatures (\cite{Graham95}).  These points suggest
that the filament segment seen is a nonradiative piece of the main blast wave
not obviously related to the other emission in the area.  The relative
faintness and lack of definition compared to other nonradiative filaments
suggests that it is not quite parallel to our line of sight.

Meanwhile, the densest material in the shocked cloud tip has cooled enough to
emit in ionic species like [\ion{O}{3}] (Figure~7b) and \ion{C}{4}.  Gas
stripping resulting from instabilities in the fluid flow along the edges of the
cloud is seen as 'windblown streamers' on the north and south as well as
diffuse emission (because of a less favorable viewing angle) to the east.  The
B5 image shows great detail of the cloud shock and closely resemble the
[\ion{O}{3}] filaments, but with an added ``tail'' extending to the southeast.
The main body of the cloud shock as viewed in B5 is likely composed of
\ion{C}{4} and \ion{O}{3}] emission while the ``tail'' may be an example of a
slow shock in a neutral medium and have an enhanced two-photon flux
(\cite{Dopita82}).  The shock velocity in the cloud is quoted by Fesen et al.
as $<$60 \vel\ though this is based on the identification of the western
segment as a reverse shock.  Given the bright [\ion{O}{3}] and B5 emission in
the cloud shock, it seems more likely that the cloud shock is similar to other
structures to the north where shock velocities are thought to be more nearly
140 \vel.

There is a general increase in signal in the northern half of the SE FUV field
(Figure~7c).  It is unclear whether this is primarily due to the background
``measles'' noted in \S2 or if this represents diffuse, hot gas emitting
\ion{C}{4} as is seen in the halo around the central knot of XA.  There is very
faint emission seen in both \Ha\ and [\ion{O}{3}] in the area which could
represent a region of more nearly face-on emitting gas.

\section{Concluding Remarks}

The UIT B5 band, although broader than ideal for SNR observations, provides a
unique FUV spectral window.  Under some conditions, the B5 bandpass provides
images of radiative filaments overrun by very high-speed shocks.  Under other
conditions, B5 observes nonradiative filaments at the extreme front edge of SNR
blast waves.  Combined with other image and spectral data, the B5 band can
provide unique insights into complex, difficult-to-model shock phenomena such
as \ion{C}{4} resonance scattering and shock completeness.

In nonradiative filaments, B5 flux comes from a mixture of \ion{C}{4}  as it
ionizes up and two-photon emission from preshock neutral hydrogen.  In general,
nonradiative filament morphology is very similar in B5 and \Ha, implying that
two-photon emission, originating in the same regions as \Ha, is the primary
contributor to the B5 images.  One unique capability of B5 imaging is its
ability to capture nonradiative shocks in ionized media.  We see one example of
such in Figure~6c where a nonradiative shock is faintly seen in \ion{C}{4} and
\ion{He}{2}.

Radiative filaments usually show good correlation between B5 and [\ion{O}{3}]
morphology, suggesting that B5 flux arises in ions with similar excitation
energies such as \ion{C}{4}.  Existing models for simple, complete shocks
indicate the same origin.

However, existing FUV spectra complicate this picture, indicating that these
regions should be dominated by two-photon flux which we would expect to follow
more closely the \Ha\ morphology.  Observational selection restricts detailed
spectral information to only the very brightest knots and filaments.
Presumably, these bright regions also suffer the greatest resonance scattering
in \ion{C}{4}$\lambda$1550, decreasing its observed flux; in fact, DBR found
unexpectedly strong resonance scattering even away from the bright filaments
and knots.  Despite this, morphological similarities between B5 and
[\ion{O}{3}] in radiative filaments strongly suggest that, at least away from
the brightest filaments and cloud cores, B5 flux is dominated by \ion{C}{4}.

\paragraph{Acknowledgements}

The authors wish to thank John Raymond for valuable discussions and the use of
unpublished HUT data.  We would also like to thank an anonymous referee for
several valuable suggestions including using FUV images to trace nonradiative
filaments through ionized regions.  Funding for the UIT project has been
through the Spacelab Office at NASA headquarters under project number 440-551.

\clearpage

\begin{figure}
\centerline{\psfig{figure=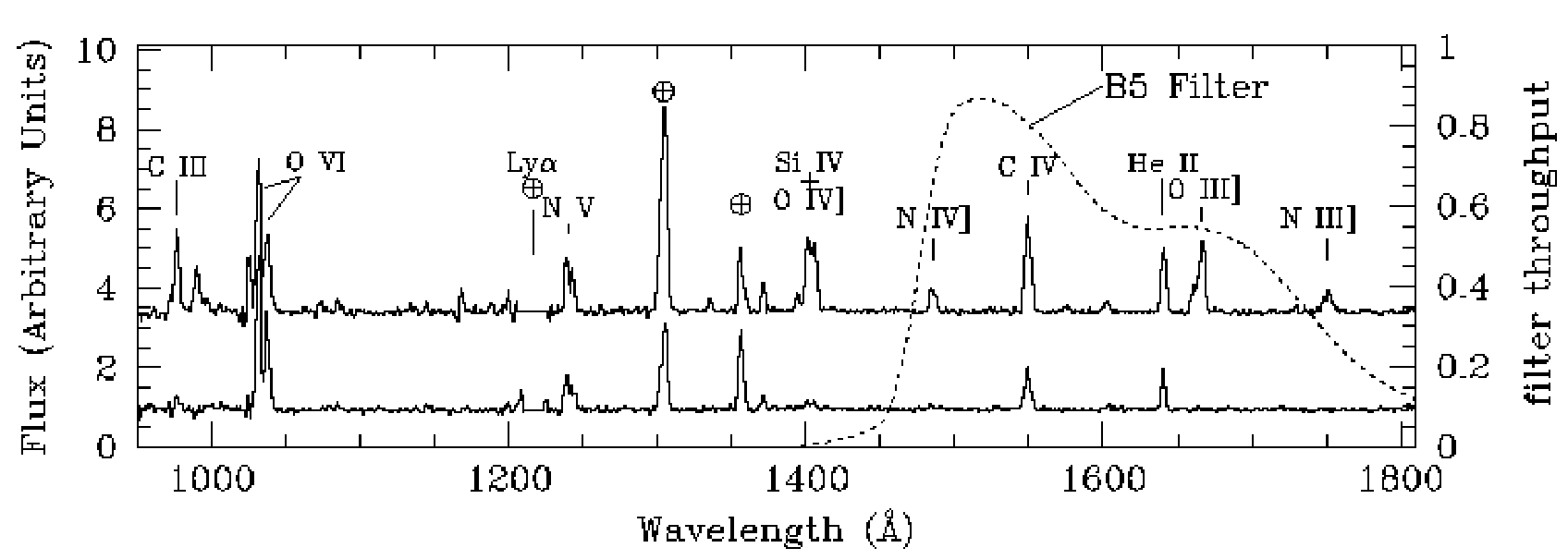,width=6in}}
\figcaption[]{\protect\small{Two typical UV spectra of SNR filaments.  The top, a
radiative filament (Blair et al. 1991), shows lines of many different ions.
The bottom, a nonradiative filament (Long et al. 1992), shows lines of only the
highest ionization species.  The dashed curve superimposed on the two spectra
represents the throughput of UIT's B5 filter as a function of wavelength.}}
\end{figure}

\begin{figure}
\centerline{\psfig{figure=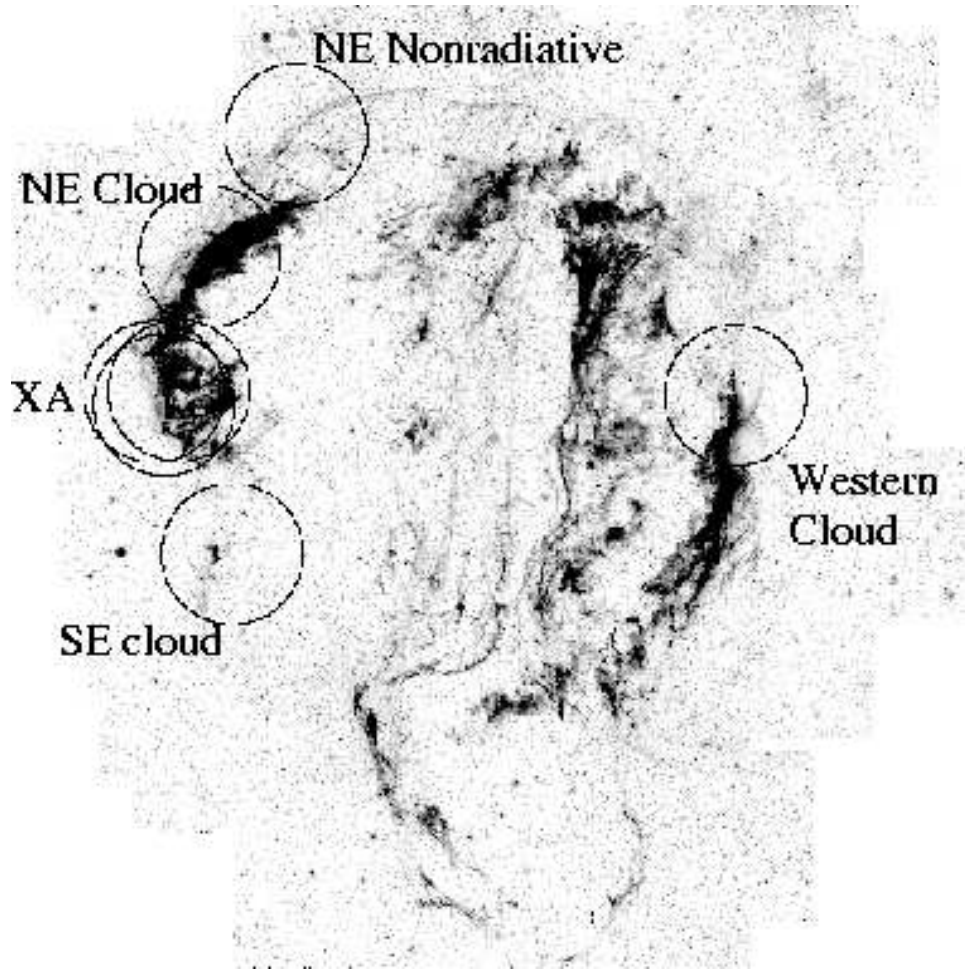,width=5in}}
\figcaption[]{\protect\small{An \Ha\ mosaic of the entire Cygnus Loop (courtesy
Levenson et al. 1998) with the 40\arcmin\ UIT fields superimposed and
labeled.}}
\end{figure}

\begin{figure}
\figcaption[]{\protect\small{The Western Cloud as viewed in a) \Ha, b)
[\protect\ion{O}{3}]$\lambda$5007, c) B5, and d) the ROSAT High Resolution Imager
(HRI). All bands show a bright north-south radiative complex similar to that
seen in the Northeast Cloud (Figure~5) but with apparently simpler geometry.
Two bright, parallel filaments suggest two points of tangency to our line of
sight.  In \Ha\ (a) and B5 (c) we also see a nonradiative filament diverging to
the northwest of the bright radiative region.  X-rays are seen behind this
filament in (d).  A reverse shock generates higher temperatures and brighter
X-ray emission at the radiative region.  With the exception of the nonradiative
filament, the B5 and the [\protect\ion{O}{3}] (b) show a high degree of correlation,
suggesting origin of the B5 flux in high-excitation ionic species.  Each field
is 40\arcmin\ across.  For Figures 3-7, both optical fields have been
median-filtered with a 3-pixel (15 \arcsec) box.  The HRI field has been
smoothed with a 3-pixel gaussian.  All fields are aligned and oriented with
north at the top and east to the left.  All image intensities are displayed
logarithmically. (Please see attached file 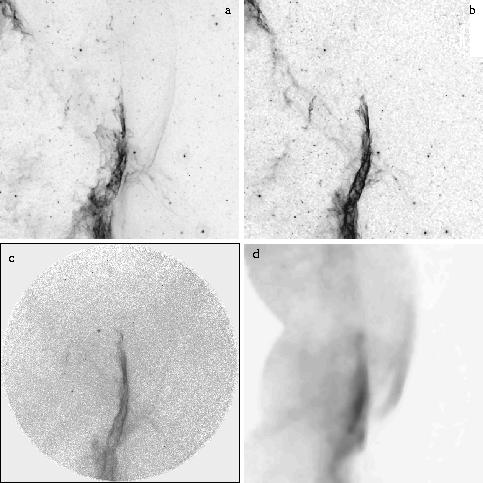 or \url\ for
full-resolution image.)}}

\figcaption[]{\protect\small{The Northeast Nonradiative Region, containing the classic
nonradiative filaments, viewed as in Figure~3.  The SN blast wave propagates
through the atomic shell at v$\sim$400 \vel. Thin filamentary emission arises
from the preshock neutral fraction as it heats up, and is seen in \Ha\ (a).
Little or no emission is seen from this filament in [\protect\ion{O}{3}] (b).  The
shock is visible in B5 (c) through both two-photon processes (closely linked to
\protect\Ha\ emission) and to a lesser extent through high-ionization species--in this
case \protect\ion{C}{4}.  The ROSAT HRI image (d) shows the $\sim10^{6}$K
X-ray-emitting post-shock gas in a band behind the shock front.  (Please see
attached file 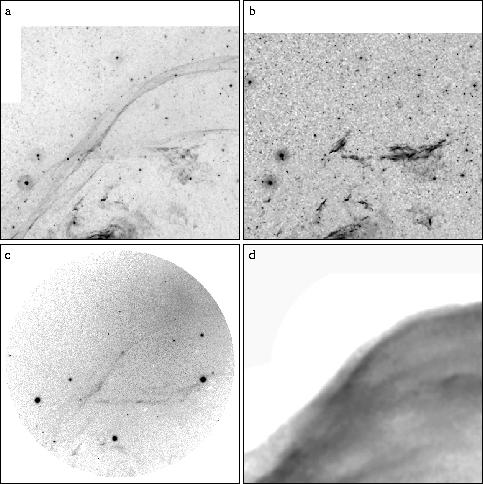 or \url\ for full-resolution image.)}}

\figcaption[]{\protect\small{The Northeast Cloud as viewed in Figure~3.  \protect\Ha\ (a) shows
smooth nonradiative filaments to the east of a more complex mass of radiative
filaments. [\protect\ion{O}{3}] (b) shows a radiative filament structure complicated by
line-of-sight coincidence of several emitting regions.  The short B5 exposure
(c) shows radiative structures well but is not deep enough to show the
nonradiative filaments.  The ROSAT image (d) shows that flux from hot gas
behind the blast wave is considerably enhanced by the strongly decelerating
shocks in the denser radiative region.  (Please see attached file 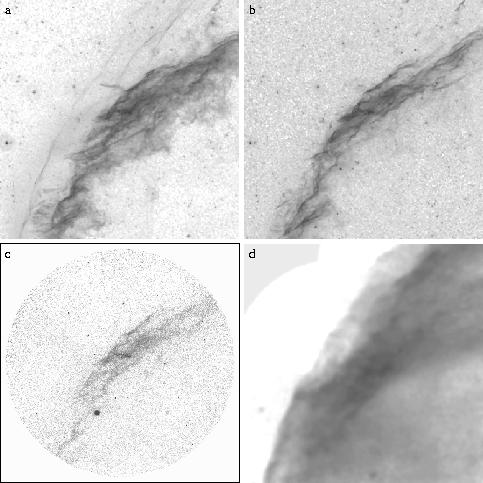
or \url\ for full-resolution image.)}}

\figcaption[]{\protect\small{The XA region as viewed in Figure~3. This region displays
a complex region of cloud-shock interactions, including dense, bright filaments
whose \protect\ion{C}{4} emission is apparently strongly affected by resonance
scattering and a range of shock completeness (see Figures 9c and 10).  (Please
see attached file 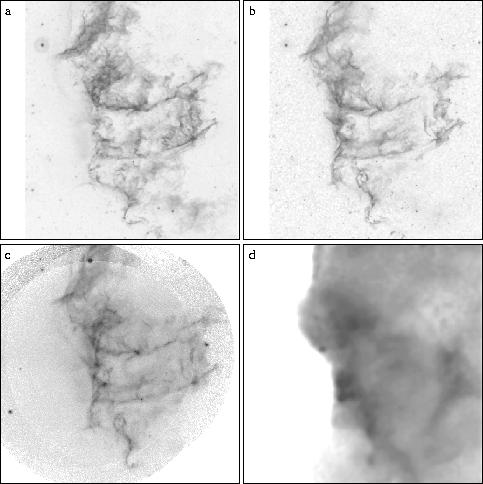 or \url\ for full-resolution image.)}}

\figcaption[]{\protect\small{The South East Cloud viewed as in Figure~3.  This small
patch of emission is likely the tip of a much larger cloud early in the stages
of shock interaction.  B5 morphology (c) closely matches both \protect\Ha\ (a) and
[\protect\ion{O}{3}] (b).  The primary differences are faint features common only to B5
and \Ha: the ``tail'' extending south of the cloud, and  faint diffuse
material--nonradiative blast wave filaments--in the southern half of the field.
 X-rays (d) show a region of emission much larger than the optical/UV cloud
with a slight flux depression or ``hole'' in the center.  (Please see attached
file 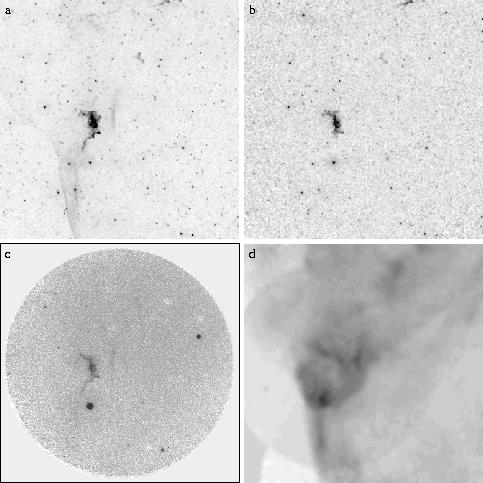 or \url\ for full-resolution image.)}}

\figcaption[]{\protect\small{Three-color images of the Western Cloud (a), the Northeast
Cloud (b), the XA Region (c) and the Southeast Cloud (d).  \protect\Ha\ is in red, B5
emission in green and X-rays in blue.  All intensities are displayed
logarithmically and colors have been adjusted to best show spatial
relationships. (Please see attached file 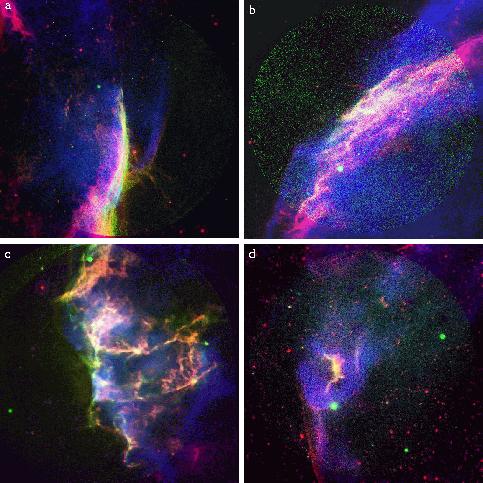 or \url\ for
full-resolution image.)}}
\end{figure}

\begin{figure}
\centerline{\psfig{figure=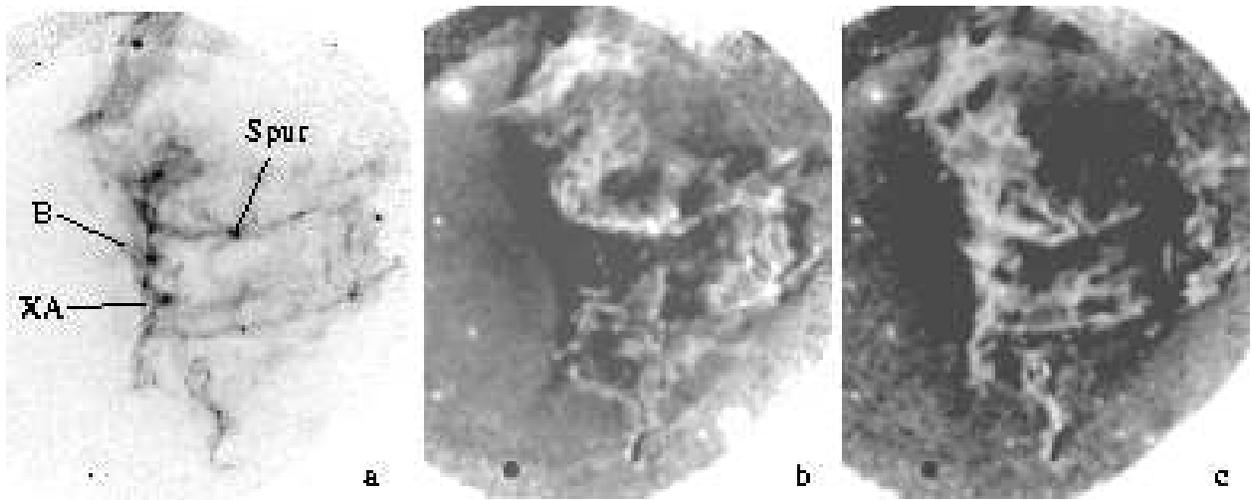,width=6.5in}}
\figcaption[]{\protect\small{The effects of completeness and resonance scattering. a)
UIT B5 image of the XA region.  b) A ``completeness map'' generated by taking
the ratio of B5 to \protect\Ha. Light areas represent more complete cooling while dark
areas are less complete.  Complete shocks emit predominantly two-photon
emission in the B5 band while incomplete regions tend toward higher \protect\ion{C}{4}
contributions.  c) ``Saturation map'' generated from B5 and [\protect\ion{O}{3}].  Dark
regions show areas of higher resonance scattering of \protect\ion{C}{4}.}}
\end{figure}

\begin{figure}
\centerline{\psfig{figure=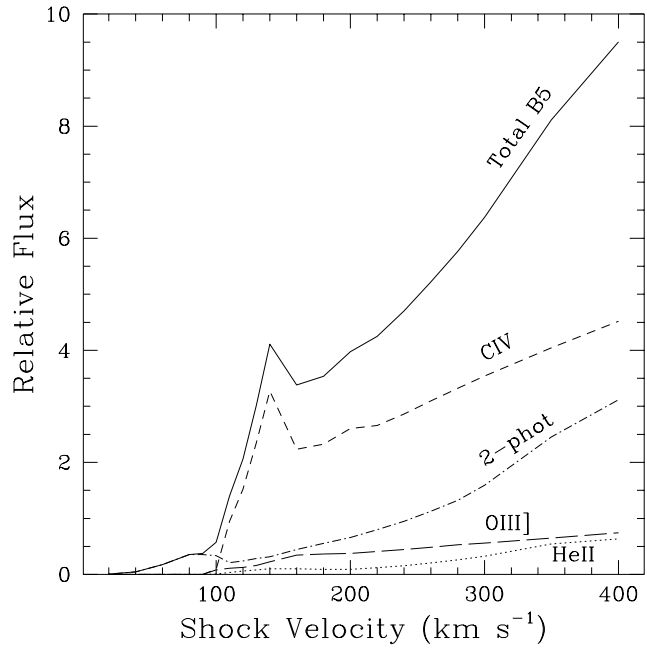,width=3.5in}}
\figcaption[]{\protect\small{B5 flux as a function of velocity for the shock models of
Hartigan, Raymond, \& Hartmann (1987).  The various dashed lines show the flux
from selected ionic species multiplied by the B5 filter throughput, as a
function of radiative shock velocity.  The solid line is the total B5 flux
calculated as a sum of 0.80$\times$\protect\ion{C}{4}$\lambda$1550,
0.55$\times$\protect\ion{He}{2}$\lambda$1640, 0.54$\times$\protect\ion{O}{3}]$\lambda$1665, and
0.15$\times$two-photon emission.  These models hold only for complete,
single-velocity shocks and do not take account of resonance scattering or other
complications.  In these models \protect\ion{C}{4} emission dominates the B5 bandpass
at all but the lowest velocities.}}\label{models}
\end{figure}

\end{document}